\documentclass[final,5p,times,twocolumn,authoryear]{elsarticle}

\usepackage{amssymb}
\usepackage{amsmath}
\usepackage{lipsum}

\usepackage{mathrsfs}

\journal{Mod. Phys Lett. A}

\begin{document}

\begin{frontmatter}




\title{A rigid spherical shell enclosing a degenerate wormhole}
\author{Juri Dimaschko}
\address{Technische Hochschule Lübeck, Mönkhofer Weg 239, 23562 Lübeck, Germany}

\begin{abstract}
A static configuration consisting of a rigid spherical shell enclosing a degenerate Schwarzschild–Klinkhamer wormhole is investigated in general relativity. The shell is analyzed using the Israel junction conditions, while the total ADM mass is assumed to remain fixed. For an ordinary rigid shell in Schwarzschild spacetime, the proper mass is given by the Brown–York relation. It is shown that replacing the flat interior by a degenerate vacuum wormhole changes the junction conditions in such a way that the proper mass of the shell vanishes, whereas the ADM mass and the exterior Schwarzschild geometry remain unchanged. The resulting configuration is therefore entirely characterized by the wormhole geometry as the carrier of the gravitational field. The physical interpretation of this result is discussed, together with its relation to the previously established collapse dynamics of degenerate wormholes. 
\end{abstract}




\end{frontmatter}
\section{Introduction}

Infinitely thin spherical shells provide one of the simplest and most widely used models for studying the interaction between matter and spacetime geometry in general relativity. Their mechanical properties are determined by the Israel junction conditions, which relate the surface energy density and pressure directly to the geometry on both sides of the shell without specifying a microscopic equation of state. This approach has found numerous applications ranging from gravitational collapse to thin-shell wormholes and gravastars \citep{israel_1966,israel_1967,visser,mazur}. In static configurations, the Brown–York formalism further provides a natural relation between the proper mass of the shell, its radius, and the total ADM mass of spacetime \citep{brown}. 

Recent studies have identified a new class of spherically symmetric vacuum solutions describing degenerate Schwarzschild–Klinkhamer wormholes \citep{klinkhamer,wang,dimaschko_2026b}. Unlike thin-shell wormholes, these configurations contain neither a material shell nor exotic matter at the throat. Instead, they represent self-consistent vacuum geometries whose ADM mass is associated with spacetime geometry alone. 

The existence of such matter-free geometries naturally raises the question of how they interact with ordinary matter. To answer this question, one needs a material system whose properties are completely determined by the surrounding geometry. The simplest configuration of this kind that preserves spherical symmetry is a rigid spherical shell that surrounds a wormhole. Since the shell is completely described by the Israel junction conditions, it provides a natural probe of the interaction between ordinary matter and a matter-free gravitational configuration. 

In the present work, we consider a rigid spherical shell enclosing a degenerate Schwarzschild–Klinkhamer wormhole and investigate the physical properties of the resulting configuration within the thin-shell formalism. We first analyze the corresponding rigid shell in an ordinary Schwarzschild spacetime, obtaining the standard Brown–York relation between the shell proper mass and the ADM mass. We then replace the flat interior with a degenerate matter-free wormhole and examine how this modification affects the Israel junction conditions and the resulting shell properties. Finally, we discuss the physical implications of the obtained configuration and its relation to the previously established dynamics of degenerate wormhole throats.

The paper is organized as follows. Section 2 briefly summarizes the properties of the degenerate Schwarzschild–Klinkhamer wormhole relevant for the present work. Section 3 establishes the reference solution for a rigid spherical shell in Schwarzschild spacetime. Section 4 considers the corresponding shell enclosing a degenerate wormhole and compares the resulting junction conditions. The physical implications of the configuration obtained are discussed in Section 5, and then Section 6 concludes.

\section{Degenerate Schwarzschild-Klinkhamer wormhole}

\textit{In this section, we briefly describe the structure and properties of the degenerate Schwarzschild-Klinkhamer wormhole.}

The simplest example of a degenerate wormhole is the spherically symmetric Schwarzschild-Klinkhamer wormhole. Its metric has the form \citep{wang}\footnote{The relativistic system of units is used, in which the speed of light \(c\) and the gravitational constant \(G\) are equal to 1, as well as the usual spherical coordinates\((r,\theta,\varphi)\),  \(d\Omega^{2} = d\theta^{2} + \sin^{2}\theta \, d\varphi^{2}\) is an element of solid angle, \(t\) is time.}
\begin{equation}
    ds^2 = \left(1 - \frac{2M}{r}\right) dt^2 - \left(1 - \frac{2M}{r}\right)^{-1} \frac{l^2}{r^2} dl^2 - r^2 d\Omega^2 ,
\end{equation}where
\begin{equation}
    r = \sqrt{l^{2} + a^{2}}
\end{equation}is the usual radial coordinate \(r\), expressed as a function of the two-sheeted coordinate \(l \in (-\infty, \infty)\), and the parameter \(a\) has the meaning of the throat radius. This metric is degenerate, since its determinant \(g\) vanishes at \(l=0\). The degeneracy of the metric causes no problems due to the possibility of a tetrad description when the connections and curvatures do not contain \(g\) in the denominators \citep{horowitz}.

Along with the two-sheeted form (6), the metric can also be represented in ordinary one-sheeted coordinates with the radial coordinate \(r\) subject to the constraint \(r \ge a\):
\begin{equation}
    ds^{2}
    =
    \begin{cases}
        (1 - 2M/r)\,dt^{2} - (1 - 2M/r)^{-1} dr^{2} - r^{2} d\Omega^{2}, & r \ge a \; (\uparrow), \\[6pt]
        (1 - 2M/r)\,dt^{2} - (1 - 2M/r)^{-1} dr^{2} - r^{2} d\Omega^{2}, & r \ge a \; (\downarrow),
    \end{cases}
\end{equation}The arrow in parentheses indicates which of the two sheets the observation point is on: the top one (↑) or the bottom one (↓). The properties of this configuration relevant for the present work areas follows:

(i) the spacetime is vacuum everywhere, including the throat region \citep{klinkhamer};

(ii) it is a vacuum configuration that generates the Schwarzschild field with ADM mass \(M\) \citep{wang}. Furthermore, this configuration creates no surface tension and no surface energy density in the throat \citep{dimaschko_2026b};

(iii) the free dynamics of the radius \(a\) describes gravitational collapse and proceeds along the radial geodesics in the Schwarzschild spacetime \citep{dimaschko_2025,dimaschko_2026c}. In this sense, the degenerate wormholes differ from thin-shell Visser wormholes, where the throat is supported by exotic matter and stable against gravitational collapse \citep{visser}.

Metric (3) describes a matter-free degenerate wormhole in a vacuum. In the next two sections, we consider the metric of a degenerate wormhole in the presence of matter in the surrounding space. To this end, a rigid spherical shell enclosing the wormhole will be introduced.  

\section{Rigid spherical shell in Schwarzschild geometry}

\textit{In this section, we begin by constructing the metric and calculating the proper mass m of a rigid spherical shell situated in an asymptotically flat Schwarzschild geometry with a given ADM mass M.}

Consider an infinitely thin spherical shell with fixed values of radius \(R\) and ADM mass \(M\). The metric can be written as
\begin{equation}
    ds^{2} =
    \begin{cases}
        \left(1 - \dfrac{2M}{r}\right) dt^{2}
        - \left(1 - \dfrac{2M}{r}\right)^{-1} dr^{2}
        - r^{2} d\Omega^{2}, & r > R, \\[6pt]
        \left(1 - \dfrac{2M}{R}\right) dt^{2}
        - dr^{2}
        - r^{2} d\Omega^{2}, & r < R,
    \end{cases}
\end{equation}  - Schwarzschild outside the sphere and flat inside. It ensures the continuity of the \(g_{tt}\) component of the metric tensor at \(r=R\), which allows one to apply Israel's boundary conditions \citep{israel_1966} and determine the pressure \(p\)  and surface energy density \(\sigma\): 
\begin{equation}
    \sigma = \frac{1}{4\pi R} \left( 1 - \sqrt{1 - \frac{2M}{R}} \right),
\end{equation}
\begin{equation}
    p = \frac{1}{8 \pi R} \left( \frac{1 - \frac{M}{R}}{\sqrt{1 - \frac{2M}{R}}} - 1 \right).
\end{equation}
The proper mass of the shell defined as  \(m = 4\pi R^2 \sigma\) is then
\begin{equation}
    m = R \left( 1 - \sqrt{1 - \frac{2M}{R}} \right).
\end{equation}This expression coincides with the Brown–York quasi-local energy of the shell–field system \citep{brown}. 

The Brown–York relation (7) expresses the proper mass \(m\) of the shell in a static state, parametrically dependent on its radius \(R\). In this regime, the rigid shell is supported by non-gravitational elastic forces that cause the internal stresses \(p\) and the surface energy density \(\sigma\), as described by (5) and (6). It is precisely this factor that accounts for the increase in the shell's proper mass m as its radius \(R\) decreases. In this sense, the limit \(R \rightarrow 2M\) corresponds to the \textit{quasi-static collapse of the rigid shell} through a controlled sequence of equilibrium states. The constancy of the ADM mass \(M\) during this process is due to the absence of gravitational wave emission in a spherically symmetric vacuum (a consequence of Birkhoff's theorem).

This situation should be contrasted with the \textit{dynamic collapse of a dust shell}, where the proper mass m constitutes an integral of motion \citep{israel_1967}. In this case, there are no internal stresses within the shell, and thus the proper mass \(m\) remains constant. The Brown–York relation is therefore associated with static configurations rather than with dynamical dust collapse. In the latter case the proper mass \(m\) is an integral of motion and is no longer determined by the Brown–York relation (7).

\section{Rigid spherical shell in Schwarzschild-Klinkhamer geometry}

\textit{In this section, we determine the metric and the proper mass m of a rigid spherical shell in the geometry of a degenerate Schwarzschild-Klinkhamer wormhole.}

For this, as in Section 2, we consider a spherical shell of radius \(R\), now containing a degenerate wormhole with a spherical throat of radius \(a<R\). As before, we assume the global ADM mass \(M\) to be fixed. In this case, the metric must be determined in three domains and has the form:
\begin{equation}
    ds^{2} =
    \begin{cases}
        \left(1 - \dfrac{2M}{r}\right) dt^{2}
        - \left(1 - \dfrac{2M}{r}\right)^{-1} dr^{2}
        - r^{2} d\Omega^{2}, \quad r \geq R \; (\uparrow),  \\[6pt]
        \left(1 - \dfrac{2\mu}{r}\right) dt^{2}
        - \left(1 - \dfrac{2\mu}{r}\right)^{-1} dr^{2}
        - r^{2} d\Omega^{2}, \quad  a \leq r \leq R \; (\uparrow),  \\[6pt]
        \left(1 - \dfrac{2M}{r}\right) dt^{2}
        - \left(1 - \dfrac{2M}{r}\right)^{-1} dr^{2}
        - r^{2} d\Omega^{2}, \quad r \geq a \; (\downarrow), 
    \end{cases}
\end{equation}Here, \(\mu\) is an auxiliary constant parameter. Identical ADM mass values on the upper and lower sheets (in both cases, \(M\)) indicate that the asymptotic regions on both sheets are consistent. The continuity of \(g_{tt}\) at \(r=R\) uniquely determines the value of \(\mu\) :
\begin{equation}
    \mu = M.
\end{equation}This fixes the metric of the entire two-sheeted configuration:
\begin{equation}
    ds^{2} =
    \begin{cases}
        \left(1 - \dfrac{2M}{r}\right) dt^{2}
        - \left(1 - \dfrac{2M}{r}\right)^{-1} dr^{2}
        - r^{2} d\Omega^{2}, \quad r \geq R \; (\uparrow),  \\[6pt]
        \left(1 - \dfrac{2M}{r}\right) dt^{2}
        - \left(1 - \dfrac{2M}{r}\right)^{-1} dr^{2}
        - r^{2} d\Omega^{2}, \quad  a \leq r \leq R \; (\uparrow),  \\[6pt]
        \left(1 - \dfrac{2M}{r}\right) dt^{2}
        - \left(1 - \dfrac{2M}{r}\right)^{-1} dr^{2}
        - r^{2} d\Omega^{2}, \quad r \geq a \; (\downarrow). 
    \end{cases}
\end{equation}It reproduces the Schwarzschild metric in the entire two-sheeted space and coincides with the metric (8) of a degenerate wormhole. In this case, the extrinsic curvature is continuous across the shell, so that the surface energy density and pressure both vanish,
\begin{equation}
    \sigma = 0, \quad p = 0
\end{equation}Hence, the proper mass of the shell is zero,
\begin{equation}
    m = 0.
\end{equation} This shows that the shell carries no proper mass in the resulting configuration. Consequently, the appearance of a degenerate wormhole inside a thin, rigid shell—provided its ADM mass \(M\) remains constant—leads to the shell's proper mass \(m\) vanishing.

A direct comparison of the metric (10) of the resulting two-sheeted configuration with the metric (3) of the degenerate Schwarzschild-Klinkhamer wormhole shows that they are now identical. Thus, the role of the carrier of a gravitational charge \(M\) moves entirely from the spherical shell (metric (4)) to the gravitational field in its current configuration (metric (10))—that is, to the matter-free wormhole (metric (3)). In the resulting configuration, the gravitational field is fully described by the geometry of the wormhole, and the ADM mass \(M\) is entirely associated with the matter-free wormhole. 

The spherical shell is no longer a source of gravitational field, and the associated internal stresses also vanish. Therefore, the wormhole throat becomes the natural dynamical object of the configuration, with its state determined by a single dynamical variable—the throat radius \(a\).

Naturally, the vanishing  proper mass of the shell does not imply the disappearance of either the shell or the gravitational field experienced by observers located on it. Their local measurements remain identical to those of an ordinary rigid shell with the same exterior Schwarzschild geometry. The difference becomes apparent only when the interior geometry is explored: the originally flat interior is replaced by a Schwarzschild spacetime that carries the same ADM mass \(M\). In this sense, the configuration illustrates the intrinsically nonlocal relation between spacetime geometry and gravitational mass in general relativity.

\section{Discussion}

\textit{The calculation presented above establishes a purely geometric property of the resulting configuration. The physical interpretation of this result, however, requires some discussion.}

The physical result obtained—that the proper mass \(m\) of a rigid spherical shell vanishes upon the appearance of a wormhole within it—is a direct consequence of Einstein's equations and the conservation of the ADM mass \(M\). In the absence of an equation of state for the matter, this result implies nothing about the specific mechanism of this transition.

At the same time, relations (5) and (6), which determine the state of the rigid spherical shell, satisfy the null energy condition and therefore correspond to an ordinary material configuration. Since both the initial metric (3) and the resulting metric (10) satisfy Einstein's equations, the equations that maintain the rigid shell in equilibrium remain valid. Thus, the disappearance of the proper mass of the shell does not signal any inconsistency of the final configuration or any violation of the conservation laws.

Since we allow for the existence of a degenerate wormhole, the initially static configuration acquires a natural dynamical degree of freedom, namely the throat radius \(a\) – and its subsequent evolution is of interest.

Recent studies  have shown that the dynamics of the throat of a degenerate Schwarzschild-Klinkhamer wormhole is described by the radial geodesics of test particles in Schwarzschild geometry. This leads to gravitational collapse into a non-traversable Einstein-Rosen wormhole, for which the throat radius equals the gravitational radius \(a=2M\).

Applied to the present shell + wormhole configuration, this can be naturally extended to the possibility of gravitational collapse of the degenerate wormhole that has formed inside the rigid static shell and assumed the role of the gravitational field source. Indeed, since the resulting metric (10) is identical to the wormhole metric (3), their dynamics are also identical and are described by the radial geodesics of test particles in Schwarzschild geometry. We will refer to this dynamic configuration as the internal collapse model. This model differs fundamentally from the ordinary model of gravitational collapse, in which the collapsing object is the shell itself. An example of ordinary collapse is the collapse of a thin dust shell \citep{israel_1967}.
\begin{figure} [t!]
 \includegraphics[scale=0.52]{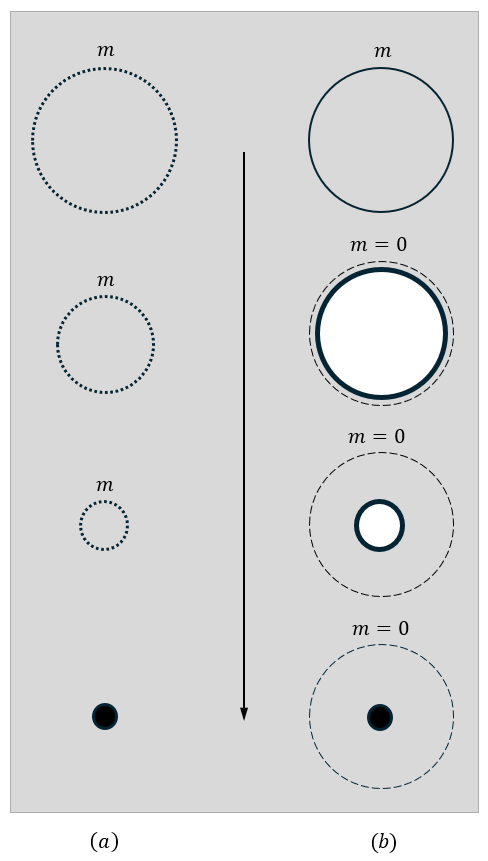}
 \centering
  \caption{Two different models of gravitational collapse: (a) ordinary collapse of a dust shell in the one-sheeted Schwarzschild spacetime; (b) internal collapse of a rigid shell in the two-sheeted Schwarzschild-Klinkhamer spacetime. \newline 
  \textbf{Legend}: dotted line – dust shell; thin solid line – rigid shell before the wormhole appearance; thin dashed line – rigid shell after the wormhole appearance; thick solid line – wormhole throat; black circle – Schwarzschild black hole (a), or non-traversable Einstein-Rosen wormhole (b).}
\end{figure}

Fig. 1 compares two models of gravitational collapse:  a) ordinary dust-shell collapse according to \citep{israel_1967} and b) internal collapse of the wormhole inside a rigid shell according to \citep{dimaschko_2025,dimaschko_2026c}. The difference between them is that:

- in ordinary collapse, the collapsing object is a material shell, whereas in internal collapse, it is a degenerate wormhole;

- ordinary collapse occurs in the one-sheeted Schwarzschild spacetime, whereas internal collapse takes place in the two-sheeted Schwarzschild-Klinkhamer spacetime; 

- in ordinary collapse, the proper mass \(m\) of the material shell is conserved, whereas in internal collapse the proper mass \(m\) of the material shell drops to zero.

Thus, the presence of a degenerate wormhole indicates a new channel of gravitational collapse, which we call internal collapse and describe within the present model. In this regard, the following should be noted:

1) On the one hand, the model of internal collapse does not provide a specific mechanism for the formation of a degenerate wormhole. Consequently, it should be regarded as an open framework rather than a complete dynamical scenario.

2) On the other hand, the model places no lower bound on the wormhole throat radius \(a\). The obtained result \(m=0\) therefore does not depend on a particular choice of \(a\), but follows from the geometric structure of the configuration itself. In this sense, the possibility of the vanishing shell proper mass is a robust consequence of the model.

The present work therefore leaves substantial room for further investigation. Among the most important open questions are the physical mechanism of wormhole formation, the stability of the resulting configurations, the role of non-spherical perturbations, and the possible influence of quantum effects.

\section{Conclusions}

In this work, we have investigated a static configuration consisting of a rigid spherical shell enclosing a degenerate Schwarzschild–Klinkhamer wormhole within the thin-shell formalism. For an ordinary shell in Schwarzschild spacetime, the proper mass is given by the standard Brown–York relation. Replacing the flat interior by a degenerate matter-free wormhole changes the Israel junction conditions in such a way that the shell proper mass vanishes while the ADM mass and the exterior Schwarzschild geometry remain unchanged.

The obtained result shows that, within the present model, the gravitational field of the resulting configuration is completely characterized by the matter-free wormhole geometry. This provides the simplest example of the interaction between ordinary matter and a matter-free vacuum geometry described within two complementary frameworks: the Israel thin-shell formalism for the material shell and the first-order description of the degenerate wormhole established in previous work \citep{dimaschko_2026b}.

The resulting configuration also provides a natural starting point for studying the subsequent collapse dynamics of degenerate wormholes inside rigid shells. The physical mechanism of wormhole formation, the stability of such configurations, and the influence of non-spherical and quantum effects remain subject for future investigation.

\section*{Acknowledgments}

I thank F.R. Klinkhamer for useful comments and stimulating discussions.

\section*{Declaration of Competing Interest}

The author declares that he has no known competing financial interests or personal relationships that could influence the content of this paper.

\section*{Data availability}

No data was used for the research described in the article.

\bibliographystyle{elsarticle-harv} 
\bibliography{main}






\end{document}